\documentclass[conference]{IEEEtran}

\usepackage{amsmath,amssymb,amsfonts}
\usepackage{graphicx}
\usepackage{booktabs}
\usepackage{multirow}
\usepackage{xcolor}
\usepackage{tikz}
\usepackage{pgfplots}
\usepackage{standalone}
\usepackage{url}
\usepackage{cite}
\usepackage[hidelinks]{hyperref}
\hypersetup{pdftitle={Channel-Token Attention for Reliable Dynamic Spectrum Access under Bursty Primary-User Traffic},pdfauthor={Krishna Acharya, Dinanath Padhya, Utsab Dahal, Ashish Kandel, Binod Sapkota},%
pdfkeywords={cognitive radio, dynamic spectrum access, reinforcement learning, resource allocation, Internet of Things, quality of service}}
\pgfplotsset{compat=1.17}
\usetikzlibrary{shapes,arrows,positioning,fit,backgrounds,calc,%
  decorations.pathreplacing,patterns,shadows,matrix}

\definecolor{urllc}{RGB}{231,76,60}
\definecolor{mmtc}{RGB}{46,204,113}
\definecolor{embb}{RGB}{52,152,219}
\definecolor{idle}{RGB}{236,240,241}
\definecolor{ppocolor}{RGB}{241,196,15}
\definecolor{attention}{RGB}{155,89,182}
\definecolor{sdnblue}{RGB}{52,73,94}
\definecolor{phygreen}{RGB}{26,188,156}
\definecolor{encoder}{RGB}{52,152,219}

\newcommand{\NumSeeds}{5}
\newcommand{\NumChannels}{20}
\newcommand{\NumDevices}{60}
\newcommand{\NumSU}{4}
\newcommand{\HistoryLen}{8}
\newcommand{\EpisodeLen}{200}
\newcommand{\NEvalSteps}{10{,}000}
\newcommand{\TrainSteps}{500{,}000}
\newcommand{\BCSteps}{20{,}000}

\newcommand{\AttnPeakNormal}{0.0999}

\newcommand{\AttnPeakExtreme}{0.1417}

\newcommand{\TacanVsMlpPP}{6.82}

\newcommand{\TacanParams}{485{,}892}

\newcommand{\ObsDim}{328}

\newcommand{\SdnDecide}{4.943}

\newcommand{\SdnTotal}{5.01}
\newcommand{\SdnTotalPtile}{17.27}

\newcommand{\AblNoModDrop}{2.89}
\newcommand{\AblNoHistDrop}{4.77}
\newcommand{\AmcDestroyDrop}{0.35}

\newcommand{\PacketGreedyAccess}{89.94}

\newcommand{\PacketGreedyDeliveryRate}{30.11}

\newcommand{\PacketGreedyDelayMean}{1.208}

\newcommand{\PacketGreedyCondGap}{9.69}

\newcommand{\PacketGreedyDeliveryGap}{0.97}

\newcommand{\PacketGreedyDelayGap}{0.312}

\newcommand{\PacketGreedyAccessNormal}{93.28}

\newcommand{\PacketGreedyAccessExtreme}{81.29}

\newcommand{\PacketMlpAccess}{83.53}

\newcommand{\PacketMlpDelayMean}{1.359}

\newcommand{\PacketTacanAssign}{92.93}

\newcommand{\PacketTacanAccess}{92.53}
\newcommand{\PacketTacanAccessStd}{0.47}
\newcommand{\PacketTacanDeliveryRate}{30.12}

\newcommand{\PacketTacanDelayMean}{1.123}

\newcommand{\PacketTacanCondGap}{3.15}

\newcommand{\PacketTacanDeliveryGap}{0.98}

\newcommand{\PacketTacanDelayGap}{0.066}

\newcommand{\PacketTacanEmptyShare}{67.59}

\newcommand{\PacketTacanAccessNormal}{93.85}

\newcommand{\PacketTacanAccessExtreme}{88.96}

\newcommand{\PacketTacanVsGreedyAccessPP}{2.59}
\newcommand{\PacketTacanVsGreedyAccessPPCILow}{1.89}
\newcommand{\PacketTacanVsGreedyAccessPPCIHigh}{3.29}
\newcommand{\PacketTacanVsGreedyAccessPPWins}{5/5}
\newcommand{\PacketTacanVsGreedyAccessPPSignP}{0.0625}

\newcommand{\PacketTacanVsGreedyAccessNormalPP}{0.57}

\newcommand{\PacketTacanVsGreedyAccessExtremePP}{7.67}
\newcommand{\PacketTacanVsMlpAccessPP}{9.00}

\newcommand{\PacketTacanVsMlpDelayReduction}{0.24}

\newcommand{\PacketGreedyUrllcAccess}{89.72}

\newcommand{\PacketGreedyUrllcDeadline}{0.16}

\newcommand{\PacketTacanUrllcAccess}{92.56}

\newcommand{\PacketTacanUrllcDeadline}{0.00}

\newcommand{\PacketGreedyMmtcAccess}{90.01}

\newcommand{\PacketTacanMmtcAccess}{92.53}

\newcommand{\PacketGreedyEmbbAccess}{89.99}

\newcommand{\PacketTacanEmbbAccess}{92.50}

\begin{document}

\pagestyle{plain}

\title{Channel-Token Attention for Reliable Dynamic Spectrum Access under Bursty Primary-User Traffic}

\author{%
\IEEEauthorblockN{Krishna Acharya, Dinanath Padhya, Utsab Dahal, Ashish Kandel, and Binod Sapkota}
\IEEEauthorblockA{Department of Electronics and Computer Engineering, Thapathali Campus\\
Institute of Engineering, Tribhuvan University, Kathmandu 44600, Nepal\\
Equal contribution: Krishna Acharya, Dinanath Padhya. Supervisor: Binod Sapkota.}
}

\maketitle

\begin{abstract}
Dynamic spectrum access must coordinate secondary users under bursty
    primary-user activity while preserving packet reliability and delay. We
    present TACAN, a centralized policy that represents each channel as a token
    containing occupancy history and automatic-modulation-classification entropy;
    a context token supplies queue class, delay and user identity. A Transformer
    encoder is warm-started from an occupancy-greedy policy and refined with
    proximal policy optimization. The frozen policies were trained to maintain a
    channel assignment in every slot, including when queues were empty. We
    therefore replay them on held-out trajectories and distinguish standby
    assignment success from packet-present access and packet delivery. In a
    \NumChannels{}-channel network with \NumDevices{} primary devices and
    \NumSU{} secondary users, TACAN achieves
    \PacketTacanAccess{}\%~$\pm$~\PacketTacanAccessStd{} packet-present access
    success, compared with \PacketGreedyAccess{}\% for Greedy and
    \PacketMlpAccess{}\% for PPO+MLP. Its paired gain over Greedy is
    \PacketTacanVsGreedyAccessPP{} points (parametric 95\% CI
    \PacketTacanVsGreedyAccessPPCILow{}--\PacketTacanVsGreedyAccessPPCIHigh{}),
    with wins in all five seeds; the exact two-sided sign-test value is
    \PacketTacanVsGreedyAccessPPSignP{}. The gain rises from
    \PacketTacanVsGreedyAccessNormalPP{} points at normal primary-user load to
    \PacketTacanVsGreedyAccessExtremePP{} points at extreme load. TACAN also
    reduces mean delivery delay from \PacketGreedyDelayMean{} to
    \PacketTacanDelayMean{} slots and the conditional user-reliability gap from
    \PacketGreedyCondGap{} to \PacketTacanCondGap{} points. Delivered packets per
    SU-slot remain arrival-limited (\PacketTacanDeliveryRate{}\% versus
    \PacketGreedyDeliveryRate{}\%), so no packet-throughput gain is claimed.
\end{abstract}


\section{Introduction}
\label{sec:intro}
%% ====================================================================

Spectrum scarcity is the binding constraint on dense IoT deployment. Sixth
generation networks are expected to carry ultra-reliable low-latency
communication (URLLC), massive machine-type communication (mMTC) and enhanced
mobile broadband (eMBB) over shared bands~\cite{ITUR2015IMT}. Cognitive radio
lets secondary users (SUs) transmit in the gaps left by licensed primary users
(PUs), but IoT traffic is bursty, heavy tailed and class dependent, so deciding
which gap to use is a hard sequential decision problem.

Deep reinforcement learning (DRL) has become a standard tool for this
problem~\cite{Yu2019JSAC-DLMA,Wang2022Sensors-UsageDSA}. Foundational agents
usually flatten the channel history, although graph encoders and attention-based
policies now supply relational structure~\cite{Jiang2020ICNC-GNNDSA,
  Yuan2023IET-GCRL,Elfikky2025CommNet-AttnDSA,Tondwalkar2026Comsnets-AttnPPO}.
We study a representation that combines relational channel structure with a
compact uncertainty statistic derived from automatic modulation classification
(AMC). Modern AMC can recover modulation classes from raw I/Q
samples~\cite{OShea2018IEEE-AMC}; here, only normalized posterior entropy is used,
not the predicted modulation label. In the simulator, class-conditioned
posterior templates make this entropy weakly correlated with primary-user
holding time. Section~\ref{sec:ablation} measures the contribution and limits of
that synthetic feature.

TACAN addresses these limitations by tokenising the spectrum. Each channel becomes one token
whose features are its occupancy history and its per-channel AMC entropy, and a
single additional QoS context token encodes the class and the normalised
queueing delay of the packet the SU is currently trying to send. A Transformer
encoder over these $K+1$ tokens gives two capabilities that a flat encoder
cannot express cheaply. Channel tokens attend to one another, so the
representation of one band can incorporate the state of the others and compare
its persistence against the load across the spectrum.
And the QoS token attends over the channels, so the same network implements a
different effective policy depending on whether it is currently serving a
latency-critical URLLC packet or a delay-tolerant mMTC report.

The policy is warm-started by behaviour cloning an occupancy-greedy heuristic
and refined with PPO. A learnable occupancy prior is added to the policy logits.
The reward uses class-dependent access and delay weights, a PU-collision penalty
and an explicit URLLC deadline penalty. These components were trained jointly;
the current experiments do not isolate their individual effects.

We also expose the policy through a controller interface and measure the
simulation-mode Python decision path. This timing is an implementation
microbenchmark, not a live SDN deployment result.

\subsection{Contributions}
\begin{itemize}
  \item \textbf{Channel tokens with a QoS query.} Building on prior
        relational spectrum policies, we represent each channel by its temporal
        occupancy and AMC entropy and add a QoS context token that re-ranks the
        channels for the service class in flight. Complexity is
        $O(L(K{+}1)^2 d)$ rather than the $O((HKM)^2d)$ of attention over the
        flattened observation.
  \item \textbf{Packet-aware checkpoint re-evaluation.} We separate continuous
        standby assignment success from access success on slots with queued
        packets, packet delivery, delay and per-user reliability. The replay uses
        frozen policies and identical held-out trajectories.
  \item \textbf{Load-dependent reliability analysis.} We evaluate mixed and
        fixed primary-user load regimes over \NumSeeds{} independently trained
        seeds and report paired effect sizes, parametric intervals and exact sign
        tests rather than treating slots as independent samples.
\end{itemize}

%% ====================================================================
\section{Related Work}
\label{sec:related}
%% ====================================================================

\textbf{DRL for dynamic spectrum access.} The canonical formulation treats
access as a POMDP solved by a deep Q-network over an occupancy history. Wang et
al.~\cite{Wang2018TCCN-DRLDSA} established this for correlated multichannel
access and compared against the Whittle index; Naparstek and
Cohen~\cite{Naparstek2019TWC-MultiUser} extended it to distributed multi-user
access without coordination signalling; Yu et al.~\cite{Yu2019JSAC-DLMA} learned
a MAC protocol from scratch for heterogeneous coexistence. Li et
al.~\cite{Li2020TCCN-SensAggr} added joint sensing and bandwidth aggregation,
Tan et al.~\cite{Tan2022IoTJ-CoopMARL} used recurrent networks under
centralised training with decentralised execution, and Zhang et
al.~\cite{Zhang2024IoTJ-Distributed} scaled distributed access to IoT.
Wang et al.~\cite{Wang2022Sensors-UsageDSA} compress occupancy history for a
double DQN, while Chang et al.~\cite{Chang2025TWC-DynaESN} combine recurrent
state with model-generated experience under partial observability. Policy
gradient methods have since proved stronger in dense
settings: Shraa and Alauthman~\cite{Shraa2025Access} report large latency
reductions for PPO over DQN, Fasihi and Mark~\cite{Fasihi2024VTC-NRU} add
priority-aware rewards for NR-U coexistence, and Song et
al.~\cite{Song2021IoTJ-SpecMgmt} apply DRL to spectrum management under
uncertainty. Broader context appears in the surveys of Luong et
al.~\cite{Luong2019COMST-Survey}. Most of these agents use flat or recurrent
encoders. A parallel line instead imposes graph structure: Jiang et
al.~\cite{Jiang2020ICNC-GNNDSA} map traffic and interference graphs to access
decisions, Yuan et al.~\cite{Yuan2023IET-GCRL} combine graph attention with DQN
for channel and power allocation, and Li et al.~\cite{Li2024AdHoc-GNNDQN} couple
a GNN and DQN for IoT DSA. These methods establish the value of relational
inductive bias, but their nodes encode users or interference relations rather
than per-channel temporal and PHY observations.

\textbf{Attention in cognitive radio.} Attention first entered mainly through
sensing. Lan et
al.~\cite{Lan2026SciRep-AttnCSS} combine graph attention with a Transformer for
cooperative \emph{sensing}; Gao et al.~\cite{Gao2024TVT} use attention to fuse
sensing reports across agents; Zhang et al.~\cite{Zhang2024TWC-SpecTrans} apply
self-attention to wideband \emph{detection}. Policy-side attention is no longer
absent. Bai et al.~\cite{Bai2025Sensors-MHSA-MADRL} weight users in a centralised
critic; Elfikky et al.~\cite{Elfikky2025CommNet-AttnDSA} place self-attention in
a distributed optical channel-access policy; and Tondwalkar and
Kwasinski~\cite{Tondwalkar2026Comsnets-AttnPPO} add attention to both PPO actor
and critic for cognitive-radio power control. Zhao et
al.~\cite{Zhao2024ChinaComm} use a Transformer for cellular resource allocation,
where the sequences represent users and resources. TACAN therefore does not
claim attention itself as novel. Its distinction is the token content and
conditioning: each policy token denotes one radio channel and jointly carries
occupancy history and an AMC-derived feature, while a QoS query token makes the
channel ranking class-conditional.

\textbf{PHY-aware access.} O'Shea et al.~\cite{OShea2018IEEE-AMC} established
that deep AMC recovers modulation class from raw I/Q over the air, and AMC is now
a mature standalone capability on commodity SDR front ends. The literature
predominantly treats it as an end in itself: the classifier output is a report,
not a policy input. To our knowledge, prior DSA policies do not feed an
AMC-derived uncertainty signal into channel-selection reinforcement learning. TACAN closes
that loop, and Sec.~\ref{sec:ablation} measures what the resulting signal is
actually worth.

\textbf{QoS-differentiated access.} Traffic-aware DSA predates deep learning:
Liu et al.~\cite{Liu2013JSAC-TrafficAware} use estimated PU traffic to order
channel sensing. Recent learning methods include the rule-assisted,
SU-traffic-aware MARL policy of Si et al.~\cite{Si2025VTC-TrafficAware}.
Iqbal et al.~\cite{Iqbal2025Sensors-PASM}
allocate channels by priority class for vehicular IoT and Qian et
al.~\cite{Qian2021JSAC-MultiOp} share spectrum across operators for massive IoT.
Barqi et al.~\cite{Barqi2021IET-FairHetNet} and Jalil et
al.~\cite{Jalil2020MobiQuitous-FairDSA} target fairness in distributed access.
These condition on class through separate policies, per-class weights or fixed
priority rules. We condition a single shared policy through a learned QoS token
that participates in the attention mechanism, so one set of weights serves all
classes and the class identity re-ranks the channels rather than re-selecting the
network.

\textbf{SDN for cognitive radio.} Kobo et al.~\cite{Kobo2017Access-SDCRN} and
Akyildiz et al.~\cite{Akyildiz2015ComNet-SoftAir} argue for SDN as the control
plane of wireless systems. Liu et al.~\cite{Liu2022TII-SDCIoT} pair
reinforcement learning with a software-defined controller for industrial IoT
spectrum access using a flat network. We keep the architectural premise and
replace the agent, and we report measured controller loop latency rather than
assuming it is negligible.

\textbf{Traffic modelling.} Realistic IoT traffic is neither Poisson nor
saturated. El Fawal et al.~\cite{ElFawal2024ApplSci-MMPP} model machine-to-machine
heterogeneous traffic with Markov-modulated Poisson processes, Paxson and
Floyd~\cite{Paxson1995WideArea} document the failure of Poisson modelling at
scale, and Gajewski et al.~\cite{Gajewski2022Sensors-TrafficCRN} argue for
Pareto service times under burstiness. Motivated by this literature, our
implemented generator is a discrete-time ON/OFF semi-Markov process with
Bernoulli activation and clipped-Pareto busy periods. This temporal persistence
makes the current-slot genie of Sec.~\ref{sec:eval} a useful reactive reference.

\textbf{Positioning.} Attention, relational encoders, SDN integration,
QoS-aware rewards and heavy-tailed traffic all have precedents. TACAN's claimed
novelty is their specific policy representation: per-channel tokens fuse
occupancy history with AMC entropy, and a QoS query token makes one shared policy
class-conditional. Table~\ref{tab:positioning} separates that representation
from attention over users, interference graphs, sensing reports and generic
state sequences.

\begin{table*}[t]
  \centering
  \caption{Position of TACAN relative to the closest prior work. Att.\ target
    is what attention operates over; PHY denotes physical-layer features in the
    observation.}
  \label{tab:positioning}
  \setlength{\tabcolsep}{3pt}
  \resizebox{\columnwidth}{!}{%
    \begin{tabular}{llllcc}
      \toprule
      \textbf{Work}                                          & \textbf{Att.\ target} & \textbf{Where}  & \textbf{QoS}   & \textbf{PHY} & \textbf{SDN} \\
      \midrule
      \cite{Wang2018TCCN-DRLDSA,Yu2019JSAC-DLMA}             & none                  & n/a             & no             & no           & no           \\
      \cite{Naparstek2019TWC-MultiUser,Tan2022IoTJ-CoopMARL} & none                  & n/a             & no             & no           & no           \\
      \cite{Zhang2024TWC-SpecTrans,Gao2024TVT}               & sensing reports       & detector        & no             & I/Q          & no           \\
      \cite{Lan2026SciRep-AttnCSS}                           & sensor nodes          & detector        & no             & I/Q          & no           \\
      \cite{Bai2025Sensors-MHSA-MADRL}                       & users                 & critic          & no             & no           & no           \\
      \cite{Jiang2020ICNC-GNNDSA,Yuan2023IET-GCRL}           & graph neighbours      & policy          & traffic        & no           & no           \\
      \cite{Elfikky2025CommNet-AttnDSA}                      & state sequence        & policy          & no             & no           & no           \\
      \cite{Tondwalkar2026Comsnets-AttnPPO}                  & state features        & actor+critic    & no             & no           & no           \\
      \cite{Zhao2024ChinaComm}                               & user requests         & policy          & no             & no           & no           \\
      \cite{Iqbal2025Sensors-PASM}                           & none                  & n/a             & rule           & no           & no           \\
      \cite{Liu2022TII-SDCIoT}                               & none                  & n/a             & no             & no           & yes          \\
      \midrule
      \textbf{TACAN}                                         & \textbf{channels}     & \textbf{policy} & \textbf{token} & \textbf{AMC} & \textbf{yes} \\
      \bottomrule
    \end{tabular}}
\end{table*}

%% ====================================================================
\section{System Model}
\label{sec:system}
%% ====================================================================

\subsection{Network Scenario}
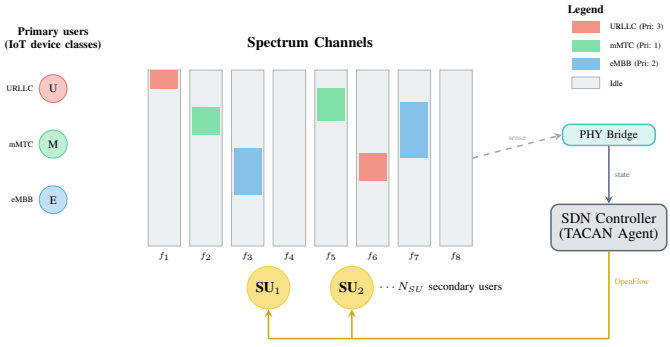
\begin{figure}[t]
  \centering
  \begin{tikzpicture}[
    scale=0.85,
    transform shape,
    device/.style={circle, draw, minimum size=0.6cm, font=\scriptsize},
    urllc/.style={device, fill=urllc!30, draw=urllc!80},
    mmtc/.style={device, fill=mmtc!30, draw=mmtc!80},
    embb/.style={device, fill=embb!30, draw=embb!80},
    su/.style={device, fill=ppocolor!50, draw=ppocolor!80, minimum size=0.8cm, font=\small\bfseries},
    channel/.style={rectangle, draw=gray!60, minimum width=0.7cm, minimum height=3.8cm, fill=idle},
    occupied/.style={rectangle, minimum width=0.6cm, minimum height=0.5cm},
    arrow/.style={->, >=stealth, thick},
    sdnbox/.style={rectangle, draw=sdnblue!80, fill=sdnblue!15, thick, rounded corners, minimum width=2.5cm, minimum height=1cm, font=\small}
  ]

  % Spectrum channels
  \foreach \i in {1,...,8} {
      \node[channel] (ch\i) at (\i*0.9, 0) {};
      \node[below, font=\tiny] at (ch\i.south) {$f_{\i}$};
    }

  % Channel label
  \node[above, font=\small\bfseries] at (4.5*0.9, 2.2) {Spectrum Channels};

  % Occupied slots visualization
  \fill[urllc!60] (0.9-0.3, 1.5) rectangle (0.9+0.3, 1.9);
  \fill[mmtc!60] (2*0.9-0.3, 0.5) rectangle (2*0.9+0.3, 1.1);
  \fill[embb!60] (3*0.9-0.3, -0.8) rectangle (3*0.9+0.3, 0.2);
  \fill[mmtc!60] (5*0.9-0.3, 0.8) rectangle (5*0.9+0.3, 1.5);
  \fill[urllc!60] (6*0.9-0.3, -0.5) rectangle (6*0.9+0.3, 0.1);
  \fill[embb!60] (7*0.9-0.3, 0.0) rectangle (7*0.9+0.3, 1.2);

  % Primary Users (IoT Devices)
  \node[urllc] (pu1) at (-1.5, 1.5) {U};
  \node[mmtc] (pu2) at (-1.5, 0.3) {M};
  \node[embb] (pu3) at (-1.5, -0.9) {E};
  \node[left, font=\tiny, align=right] at (pu1.west) {URLLC};
  \node[left, font=\tiny, align=right] at (pu2.west) {mMTC};
  \node[left, font=\tiny, align=right] at (pu3.west) {eMBB};
  \node[above, font=\scriptsize\bfseries, align=center] at (-1.5, 2.2)
    {Primary users\\(IoT device classes)};

  % Secondary Users
  \node[su] (su1) at (3.5*0.9, -2.8) {$\text{SU}_1$};
  \node[su] (su2) at (5.5*0.9, -2.8) {$\text{SU}_2$};
  \node[right, font=\scriptsize] at (su2.east) {$\cdots N_{SU}$ secondary users};

  % SDN Controller
  \node[sdnbox] (sdn) at (10.5, -1.5) {\shortstack{SDN Controller\\(TACAN Agent)}};

  % PHY Bridge
  \node[rectangle, draw=phygreen!80, fill=phygreen!15, thick, rounded corners, minimum width=2cm, font=\scriptsize] (phy) at (10.5, 0.5) {PHY Bridge};

  % Arrows
  \draw[arrow, dashed, gray!70] (ch8.east) -- (phy.west) node[midway, above, font=\tiny] {sense};
  \draw[arrow, sdnblue!80] (phy) -- (sdn) node[midway, right, font=\tiny] {state};
  % OpenFlow rules are pushed to every SU over a bus routed clear of the
  % SU row, so no arrow passes through another node.
  \coordinate (ofbus) at (10.5, -3.9);
  \draw[ppocolor!85!black, thick] (sdn.south) -- (ofbus)
    node[pos=0.35, right, font=\tiny, ppocolor!85!black] {OpenFlow};
  \draw[ppocolor!85!black, thick] (ofbus) -- (su1.south |- ofbus);
  \draw[arrow, ppocolor!85!black, thick] (su1.south |- ofbus) -- (su1.south);
  \draw[arrow, ppocolor!85!black, thick] (su2.south |- ofbus) -- (su2.south);

  % Legend
  \begin{scope}[shift={(10, 2)}]
    \node[font=\scriptsize\bfseries] at (0, 1.2) {Legend};
    \fill[urllc!60] (-0.3, 0.7) rectangle (0.3, 0.95);
    \node[right, font=\tiny] at (0.4, 0.825) {URLLC (Pri: 3)};
    \fill[mmtc!60] (-0.3, 0.3) rectangle (0.3, 0.55);
    \node[right, font=\tiny] at (0.4, 0.425) {mMTC (Pri: 1)};
    \fill[embb!60] (-0.3, -0.1) rectangle (0.3, 0.15);
    \node[right, font=\tiny] at (0.4, 0.025) {eMBB (Pri: 2)};
    \fill[idle] (-0.3, -0.5) rectangle (0.3, -0.25);
    \draw[gray!60] (-0.3, -0.5) rectangle (0.3, -0.25);
    \node[right, font=\tiny] at (0.4, -0.375) {Idle};
  \end{scope}

\end{tikzpicture}
  \caption{SDN-enabled cognitive radio network. Primary IoT devices of three
    service classes occupy channels with class-specific traffic and modulation.
    The controller runs the TACAN policy and produces channel-assignment
    recommendations for the secondary users.}
  \label{fig:network_scenario}
\end{figure}

We consider $K$ orthogonal channels $\mathcal{C}=\{1,\dots,K\}$ shared by
$N_{PU}$ primary IoT devices and $N_{SU}$ secondary users coordinated by a
centralised SDN controller (Fig.~\ref{fig:network_scenario}). Each PU belongs to
one of the three 3GPP service classes and is assigned to one channel. SUs
opportunistically transmit on channels they believe to be idle. If two SUs pick
the same channel both fail, so the controller must also avoid self-collisions.

\subsection{Traffic and PHY Model}
Each PU device follows a discrete-time ON/OFF semi-Markov process. While idle,
it activates with probability $p_{\mathrm{act}}=1-e^{-\lambda_d l}$, the
probability of at least one event in a Poisson interval at load multiplier $l$.
At activation, its busy duration is a rounded, clipped Pareto draw,
\begin{equation}
  \begin{split}
    \tau & = \operatorname{round}\big(\min\{H,\max\{L,x_m(1+Y)\}\}\big), \\
    Y    & \sim \operatorname{Pareto}(\alpha),
  \end{split}
\end{equation}
which produces persistent, heavy-tailed busy periods with endpoint clipping. Class
parameters are given in Table~\ref{tab:classes}.

Each simulated class carries a characteristic distribution over $M=10$
modulation schemes. URLLC is dominated by BPSK and QPSK, mMTC by OOK and BPSK,
and eMBB by 16QAM. At
each step the simulator samples a class-conditioned AMC posterior
$\mathbf{m}_{t,k}\in\mathbb{R}^{M}$ per channel, from which we compute the
normalised entropy
\begin{equation}
  \label{eq:modent}
  e_{t,k}=-\frac{1}{\log M}\sum_{j=1}^{M} m_{t,k,j}\log m_{t,k,j}.
\end{equation}
We use entropy because it is permutation invariant and compresses $M$ dimensions
to one, keeping the per-channel token small. It preserves posterior concentration,
not the identity of the predicted modulation. The class-conditioned templates
have different concentration and therefore make entropy class-correlated in this
simulation; validating that correlation with a real noisy AMC front end remains
outside the present evidence.

\begin{table}[t]
  \centering
  \caption{IoT traffic class parameters. Durations are in slots.}
  \label{tab:classes}
  \begin{tabular}{lccccl}
    \toprule
    \textbf{Class} & \textbf{Pri.} & $\boldsymbol{\lambda}$ & $\boldsymbol{\alpha}$
                   & \textbf{Dur.} & \textbf{Dominant mod.}                                                           \\
    \midrule
    URLLC          & 3             & 1/8                    & 0.8                   & 1 to 10 & BPSK 40\%, QPSK 50\%  \\
    mMTC           & 1             & 1/60                   & 0.5                   & 1 to 5  & OOK 35\%, BPSK 40\%   \\
    eMBB           & 2             & 1/12                   & 1.5                   & 5 to 40 & 16QAM 50\%, 8PSK 15\% \\
    \bottomrule
  \end{tabular}
\end{table}

\subsection{Decision Problem}
The task is a partially observable Markov decision process
$\langle\mathcal{S},\mathcal{A},\mathcal{P},R,\gamma\rangle$. Secondary user $i$
observes
\begin{equation}
  \label{eq:obs}
  s_{t,i}=\big[\,\mathbf{O}_{t}\;\big\|\;\mathbf{E}_{t}\;\big\|\;\mathbf{q}_i\;\big\|\;d_i\;\big\|\;\mathbf{u}_i\,\big],
\end{equation}
where $\mathbf{O}_t\in\{0,1\}^{H\times K}$ is the occupancy history over the
last $H$ slots, $\mathbf{E}_t\in[0,1]^{H\times K}$ the matching modulation
entropy history from~\eqref{eq:modent}, $\mathbf{q}_i\in\{0,1\}^{3}$ a one-hot
class indicator for the head-of-line packet, $d_i$ its normalised queueing
delay, and $\mathbf{u}_i$ a one-hot SU identifier. With $H=\HistoryLen{}$,
$K=\NumChannels{}$ and $N_{SU}=\NumSU{}$ this gives \ObsDim{} dimensions per SU.
The action $a_{t,i}\in\{1,\dots,K\}$ is a channel index; the controller emits
$N_{SU}$ actions per slot. This implementation maintains an assignment for every
SU in every slot. If a queue is empty, the assignment is a standby channel
recommendation rather than a packet transmission. Section~\ref{sec:eval}
therefore reports assignment success over all slots separately from access
success on slots where a packet is queued. The objective is the usual discounted return
$J(\pi)=\mathbb{E}_{\tau\sim\pi}\big[\sum_t\gamma^t\sum_i r_{t,i}\big]$.

%% ====================================================================
\section{The TACAN Architecture}
\label{sec:method}
%% ====================================================================

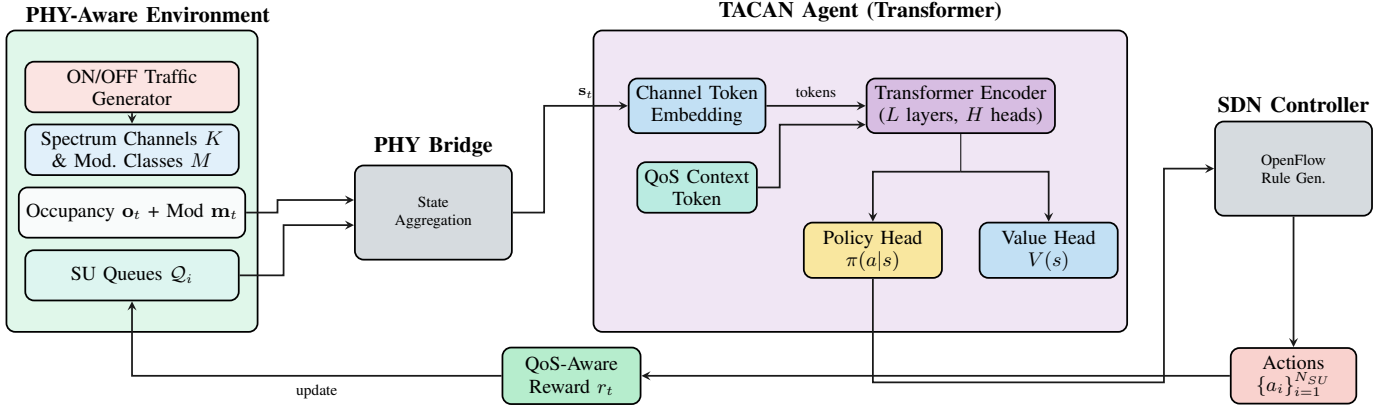
\begin{figure*}[t]
  \centering
  \begin{tikzpicture}[
    scale=1,
    transform shape,
    block/.style={rectangle, draw, thick, rounded corners, minimum height=0.8cm, font=\small, align=center},
    env/.style={rectangle, draw, thick, rounded corners, fill=mmtc!15, minimum width=4cm, minimum height=4.8cm},
    agent/.style={rectangle, draw, thick, rounded corners, fill=attention!15, minimum width=8.5cm, minimum height=4.8cm},
    sdn/.style={block, fill=sdnblue!20, minimum width=2.5cm},
    arrow/.style={->, >=stealth, thick, draw=black!90},
  ]

  % ==========================================
  % 1. BACKGROUND CONTAINERS
  % ==========================================
  % Drawn first so they sit securely behind the inner components

  % Environment Container
  \node[env] (env) at (0, 0.3) {};
  \node[above right, font=\normalsize\bfseries] at ([xshift=2mm]env.north west) {PHY-Aware Environment};

  % TACAN Agent Container
  \node[agent] (agent) at (11.6, 0.3) {};
  \node[above, font=\normalsize\bfseries] at (agent.north) {TACAN Agent (Transformer)};

  % ==========================================
  % 2. INNER BLOCKS & COMPONENTS
  % ==========================================

  % --- Environment components ---
  \node[block, fill=urllc!15, minimum width=3.4cm] (traffic) at (0, 1.8) {ON/OFF Traffic\\Generator};
  \node[block, fill=embb!15, minimum width=3.4cm] (spectrum) at (0, 0.8) {Spectrum Channels $K$\\ \& Mod. Classes $M$};
  \node[block, fill=idle!30, minimum width=3.4cm] (occ) at (0, -0.2) {Occupancy $\mathbf{o}_t$ + Mod $\mathbf{m}_t$};
  \node[block, fill=phygreen!15, minimum width=3.4cm] (queues) at (0, -1.2) {SU Queues $\mathcal{Q}_i$};

  % --- PHY Bridge ---
  \node[sdn, minimum height=1.5cm] (phy) at (4.8, -0.2) {};
  \node[above, font=\normalsize\bfseries] at (phy.north) {PHY Bridge};
  \node[font=\scriptsize, align=center] at (phy.center) {State\\Aggregation};

  % --- Agent components ---
  \node[block, fill=encoder!30] (tok_embed) at (9.0, 1.5) {Channel Token\\Embedding};
  \node[block, fill=phygreen!30] (qos_tok) at (9.0, 0.2) {QoS Context\\Token};
  \node[block, fill=attention!40, minimum width=3cm] (transformer) at (13.2, 1.5) {Transformer Encoder\\($L$ layers, $H$ heads)};
  \node[block, fill=ppocolor!40, minimum width=2.2cm] (policy) at (11.8, -0.8) {Policy Head\\$\pi(a|s)$};
  \node[block, fill=embb!30, minimum width=2.2cm] (value) at (14.6, -0.8) {Value Head\\$V(s)$};

  % --- SDN Controller & Actions ---
  \node[sdn, minimum height=1.5cm] (sdn) at (18.5, 0.5) {};
  \node[above, font=\normalsize\bfseries] at (sdn.north) {SDN Controller};
  \node[font=\scriptsize, align=center] at (sdn.center) {OpenFlow\\Rule Gen.};

  \node[block, fill=urllc!25, minimum width=2cm] (action) at (18.5, -2.8) {Actions\\$\{a_i\}_{i=1}^{N_{SU}}$};

  % --- Reward Box ---
  \node[block, fill=mmtc!35, minimum width=2.2cm, minimum height=0.9cm] (reward) at (7.0, -2.8) {QoS-Aware\\Reward $r_t$};

  % ==========================================
  % 3. PERFECTLY ROUTED ARROWS
  % ==========================================

  % Inside Environment
  \draw[arrow] (traffic) -- (spectrum);

  % Environment to PHY Bridge
  \draw[arrow] (occ.east) -- ++(0.5,0) |- ([yshift=0.2cm]phy.west);
  \draw[arrow] (queues.east) -- ++(0.7,0) |- ([yshift=-0.2cm]phy.west);

  % PHY Bridge to Agent (Notice the \mathbf{s}_t is now lowercase as in the image)
  \draw[arrow] (phy.east) -- ++(0.5,0) |- (tok_embed.west) node[pos=0.75, above, font=\scriptsize] {$\mathbf{s}_t$};

  % Inside Agent
  \draw[arrow] (tok_embed.east) -- (transformer.west) node[midway, above, font=\scriptsize] {tokens};
  \draw[arrow] (qos_tok.east) -- ++(0.35,0) |- ([yshift=-0.3cm]transformer.west);

  % Branching arrow cleanly dropping out of Transformer into the Heads
  \draw (transformer.south) -- ++(0,-0.6) coordinate (t_split);
  \draw[arrow] (t_split) -| (policy.north);
  \draw[arrow] (t_split) -| (value.north);

  % Policy to SDN: Routes strictly out of the agent block, right, up, and right again into the left face of SDN
  \draw[arrow] (policy.south) -- ++(0,-1.65) -| ([xshift=-0.8cm]sdn.west) -- (sdn.west);

  % SDN to Actions
  \draw[arrow] (sdn.south) -- (action.north);

  % Actions to Reward
  \draw[arrow] (action.west) -- (reward.east);

  % Reward back to Environment (Specifically entering the bottom of SU Queues)
  \draw[arrow] (reward.west) -| (queues.south) node[near start, below, font=\scriptsize] {update};

\end{tikzpicture}
  \caption{TACAN pipeline. The PHY bridge aggregates occupancy and AMC output
    into the observation. Each channel becomes one token; the serving SU's
    traffic class and delay become a QoS context token. A Transformer encoder
    mixes them, per-channel logits form the policy and the QoS token drives the
    value head. The controller converts assignments into rule descriptions.}
  \label{fig:system_overview}
\end{figure*}

Fig.~\ref{fig:system_overview} shows the complete pipeline. The design question
is how to turn~\eqref{eq:obs}, a flat vector, into something whose structure
matches the problem.

\subsection{Spectrum Tokenisation}
For channel $k$ we assemble a token from the three quantities the agent needs
about that band: its occupancy history, its modulation-entropy history, and its
mean occupancy, which is exactly the statistic the greedy heuristic acts on and
which therefore gives the network direct access to the baseline signal:
\begin{equation}
  \label{eq:token}
  \mathbf{x}_k=\Big[\,\mathbf{o}_{:,k}\;\big\|\;\mathbf{e}_{:,k}\;\big\|\;\tfrac{1}{H}\textstyle\sum_{h}o_{h,k}\,\Big]\in\mathbb{R}^{2H+1},
\end{equation}
\begin{equation}
  \mathbf{z}_k=\mathrm{GELU}\big(\mathrm{LN}(\mathbf{W}_{e}\mathbf{x}_k+\mathbf{b}_e)\big)\in\mathbb{R}^{d}.
\end{equation}
The QoS context of the serving SU is projected by a separate head into one
additional token $\mathbf{z}_{\mathrm{QoS}}=\mathrm{GELU}(\mathrm{LN}(\mathbf{W}_{q}[\mathbf{q}_i\|d_i]+\mathbf{b}_q))$.
Learnable positional embeddings $\mathbf{P}\in\mathbb{R}^{(K+1)\times d}$ are
added to the sequence $\mathbf{Z}=[\mathbf{z}_{\mathrm{QoS}},\mathbf{z}_1,\dots,\mathbf{z}_K]$.

Unlike a flat encoder, Eq.~\eqref{eq:token} applies one shared feature extractor
to all channels. The subsequent attention layers compare the resulting channel
representations.

\subsection{Encoder and Heads}
$L$ pre-norm Transformer blocks are applied,
\begin{align}
  \mathbf{Z}'  & =\mathbf{Z}+\mathrm{MHA}\big(\mathrm{LN}(\mathbf{Z})\big),   \\
  \mathbf{Z}'' & =\mathbf{Z}'+\mathrm{FFN}\big(\mathrm{LN}(\mathbf{Z}')\big),
\end{align}
with $N_H$ heads and a two-layer GELU feed-forward network of width $4d$.
Writing $\mathbf{H}=[\mathbf{h}_{\mathrm{QoS}},\mathbf{h}_1,\dots,\mathbf{h}_K]$
for the output, the policy takes per-channel logits through a shared linear head
and adds a learnable residual on the greedy occupancy prior,
\begin{equation}
  \label{eq:logits}
  \ell_k=\mathbf{w}_{\pi}^{\top}\mathbf{h}_k-\alpha_g\sum_{h=1}^{H}o_{h,k},
  \qquad \pi(a{=}k\mid s)=\mathrm{softmax}_k(\ell),
\end{equation}
where $\alpha_g$ is a scalar parameter learned jointly with the rest of the
network. The second term encodes the heuristic that emptier channels are better;
learning $\alpha_g$ lets the agent decide how much to trust it, and the network
output supplies the correction. The value head reads the QoS token pooled with
the channel mean, $V(s)=\mathrm{MLP}(\mathbf{h}_{\mathrm{QoS}}+\frac{1}{K}\sum_k\mathbf{h}_k)$.
An auxiliary head predicts next-slot occupancy per channel from $\mathbf{h}_k$;
its loss is added with weight $0.1$ and acts as a representation regulariser
that forces the encoder to be predictive rather than merely descriptive.

Attention over $K+1$ tokens costs $O(L(K{+}1)^2d)$ per forward pass. Applying
attention to the raw observation instead would cost $O((HKM)^2 d)$, which for
our configuration is four orders of magnitude larger.

\subsection{QoS-Differentiated Reward}
For SU $i$ serving a class-$c$ packet at step $t$,
\begin{equation}
  \label{eq:reward}
  \begin{split}
    r_{t,i}={} & w_T^{(c)}\,\mathbb{I}(\mathrm{success}_i)
    -w_D^{(c)}\min\!\Big(\tfrac{d_i}{d_{\max}},1\Big)         \\
               & -\beta\,\mathbb{I}(\mathrm{PU\ collision}_i)
    -\delta\,\mathbb{I}(c=\mathrm{URLLC},d_i>10),
  \end{split}
\end{equation}
with $\beta=1.5$, $\delta=5.0$ and $d_{\max}=50$. The executed class weights are
$(w_T,w_D)=(3.0,2.0)$ for URLLC, $(1.0,0.2)$ for mMTC and $(1.5,0.5)$ for eMBB.
Thus URLLC receives the largest success weight, a delay weight ten times that of
mMTC and an additional penalty after ten slots. The reported checkpoints use no
switching or fairness bonus; those optional implementation coefficients are zero.
During legacy training, empty queues receive a zero-valued queue context and are
treated as mMTC for reward lookup, while collision-free standby assignments still
receive the success term. The packet-aware analysis in Sec.~\ref{sec:eval}
removes these standby outcomes from packet-attempt and delivery metrics, but it
does not retrain the policies.

\subsection{Training}
Training has two stages. First the policy is behaviour cloned for
\BCSteps{} steps against the occupancy-greedy heuristic, which places it in a
competent region of policy space and avoids the long random-exploration phase in
which PPO would otherwise collide constantly. Then PPO~\cite{Schulman2017PPO}
with generalised advantage estimation optimises the clipped surrogate
\begin{equation}
  L^{\mathrm{CLIP}}(\theta)=\mathbb{E}_t\Big[\min\big(\rho_t\hat{A}_t,\;
    \mathrm{clip}(\rho_t,1{-}\epsilon,1{+}\epsilon)\hat{A}_t\big)\Big],
\end{equation}
$\rho_t=\pi_\theta(a_t|s_t)/\pi_{\theta_{\mathrm{old}}}(a_t|s_t)$, with the
value, entropy and auxiliary-prediction losses added. Transitions from all
$N_{SU}$ users share one buffer with per-minibatch advantage normalisation.
The learning rate is cosine annealed from $3\times10^{-4}$ to $10^{-5}$ after a
3\% warmup, the entropy coefficient from $0.03$ to $0.01$, and gradients are
clipped at $0.5$. TACAN and PPO+MLP share the PPO schedule, reward, warm start
and seeds. TACAN additionally has the auxiliary prediction head, the learned
greedy-prior term and 485,892 parameters, whereas the MLP has 155,413 parameters.
Their comparison therefore evaluates complete training configurations and does
not isolate attention from supervision, prior structure or capacity.

%% ====================================================================
\section{Controller Integration}
\label{sec:sdn}
%% ====================================================================

The trained policy is exposed through a three-stage controller interface. A
simulation bridge collects occupancy, AMC posteriors, history and queue state; a
decision engine builds the observation in Eq.~\eqref{eq:obs}; and a rule builder
converts channel assignments into Python dictionaries with OpenFlow-like fields.
The default polling interval is 500\,ms.

The repository also contains prototype Ryu and Mininet-WiFi adapters
inspired by~\cite{Fontes2015MininetWiFi}. They are not exercised in the reported
experiments: the live bridge uses placeholder queue state, and no OpenFlow
transport, switch installation or radio retuning is included in the latency
measurement. Section~\ref{sec:sdnlat} therefore reports only the simulation-mode
decision path.

%% ====================================================================
\section{Evaluation}
\label{sec:eval}
%% ====================================================================

\subsection{Setup}
The environment has $K=\NumChannels{}$ channels, $N_{PU}=\NumDevices{}$ primary
devices, $N_{SU}=\NumSU{}$ secondary users, history length $H=\HistoryLen{}$ and
episodes of \EpisodeLen{} slots. At reset, the load multiplier is sampled as
1.0, 1.5 or 2.5 with probabilities 0.50, 0.35 and 0.15. TACAN and PPO+MLP are
trained for \TrainSteps{} environment steps under \NumSeeds{} seeds
(42, 123, 7, 2024, 314). We replay each frozen policy for \NEvalSteps{} steps on
matched held-out seeds under mixed load and each fixed primary-user load. No
retraining is performed for the packet-aware analysis. A slot is a simulation
decision epoch without an assumed mapping to wall-clock duration; delay values
are therefore reported in slots rather than milliseconds.

We distinguish four outcomes. \emph{Assignment success} is the legacy metric:
the recommended channel is free of PU and SU contention, including standby
assignments made for empty queues. \emph{Packet-present access success} is the
fraction of queued transmission attempts that deliver a packet. \emph{Delivery
  rate} is delivered packets per SU-slot, and delivery delay is measured in slots.
Packet accounting includes synthetic reset packets, stochastic arrivals, silent
queue-capacity drops and packets censored at episode boundaries.

The comparison includes a coordinated random policy that selects distinct
channels, \emph{Greedy}, \emph{PPO+MLP}, TACAN and a \emph{current-slot genie}.
The PPO+MLP configuration shares the observation, reward, warm start, PPO
schedule and seeds, but it is smaller and lacks TACAN's auxiliary head and
greedy-prior term; it is not an attention-only ablation. DDQN is omitted because
only a cached scalar, not a matched checkpoint, is available.

\subsection{Overall Performance}
Table~\ref{tab:packet_main} separates standby assignment from packet service.
The legacy TACAN assignment-success rate is \PacketTacanAssign{}\%, but
\PacketTacanEmptyShare{}\% of its successful assignments occur while the queue
is empty. This value is therefore not packet throughput. On slots with a queued
packet, TACAN succeeds on \PacketTacanAccess{}\%~$\pm$~\PacketTacanAccessStd{} of
attempts, compared with \PacketGreedyAccess{}\% for Greedy and
\PacketMlpAccess{}\% for PPO+MLP. Delivered packets per SU-slot are nearly equal
for TACAN (\PacketTacanDeliveryRate{}\%) and Greedy
(\PacketGreedyDeliveryRate{}\%) because the fixed 30\% arrival process limits
offered traffic. We therefore claim improved access reliability and delay, not
higher packet throughput. TACAN's mean delivery delay is
\PacketTacanDelayMean{} slots, against \PacketGreedyDelayMean{} for Greedy and
\PacketMlpDelayMean{} for PPO+MLP.

% AUTO-GENERATED -- do not edit
\begin{table*}[t]
  \centering
  \caption{Packet-aware re-evaluation of the frozen policies (mean $\pm$
  standard deviation over five seeds). Assignment success includes standby
  assignments; access success is conditioned on a queued packet. Del./slot is
  delivered packets per SU-slot, and user gap is the max--min conditional-access
  gap.}
  \label{tab:packet_main}
  \resizebox{\textwidth}{!}{%
  \begin{tabular}{lcccccc}
    \toprule
    \textbf{Method} & \textbf{Assignment (\%)} &
    \textbf{Packet-present access (\%)} & \textbf{Del./slot (\%)} &
    \textbf{Mean delay} & \textbf{P95 delay} & \textbf{User gap (pp)} \\
    \midrule
        Random (coordinated) & 51.65 $\pm$ 0.60 & 50.26 $\pm$ 0.85 & 29.67 & 3.760 & 11.850 & 1.29 \\
    Greedy & 91.60 $\pm$ 0.32 & 89.94 $\pm$ 0.57 & 30.11 & 1.208 & 2.000 & 9.69 \\
    PPO+MLP & 86.11 $\pm$ 0.77 & 83.53 $\pm$ 0.98 & 30.08 & 1.359 & 3.000 & 9.75 \\
    \textbf{TACAN} & \textbf{92.93 $\pm$ 0.14} & \textbf{92.53 $\pm$ 0.47} & \textbf{30.12} & \textbf{1.123} & \textbf{2.000} & \textbf{3.15} \\
    Current-slot genie & 88.30 $\pm$ 0.41 & 88.13 $\pm$ 0.32 & 30.10 & 1.203 & 2.000 & 1.45 \\
    \bottomrule
  \end{tabular}}
  \vspace{2pt}
  \parbox{\textwidth}{\footnotesize The policies were trained under continuous
  per-slot channel assignment. These packet-aware metrics are a post hoc replay
  of the frozen checkpoints; no retraining was performed.}
\end{table*}

\subsection{Paired Uncertainty}
Table~\ref{tab:packet_comparison} reports paired differences across the five
independently trained seeds. TACAN improves packet-present access over Greedy by
\PacketTacanVsGreedyAccessPP{} points, with a parametric 95\% interval of
\PacketTacanVsGreedyAccessPPCILow{} to
\PacketTacanVsGreedyAccessPPCIHigh{} points, and wins in
\PacketTacanVsGreedyAccessPPWins{} seeds. However, five pairs are too few for a
distribution-free significance claim: the exact two-sided sign-test value is
\PacketTacanVsGreedyAccessPPSignP{}. We therefore emphasize the paired effect
size, interval and consistency rather than treating a small-sample $t$-test as
decisive.
The PPO+MLP difference is larger but remains a whole-configuration comparison,
not an attention-only effect.

% AUTO-GENERATED -- do not edit
\begin{table}[htbp]
  \centering
  \caption{Paired packet-aware differences for TACAN. Confidence intervals
  are parametric Student-$t$ intervals; the exact sign test is reported because
  only five trained seeds are available.}
  \label{tab:packet_comparison}
  \resizebox{\columnwidth}{!}{%
  \begin{tabular}{llcccc}
    \toprule
    \textbf{TACAN vs.} & \textbf{Metric} & \textbf{Mean} &
    \textbf{95\% CI} & \textbf{Wins} & \textbf{Exact $p$} \\
    \midrule
        Greedy & Access success (pp) & +2.59 & [1.89, 3.29] & 5/5 & 0.0625 \\
    Greedy & Mean-delay reduction & +0.09 & [0.06, 0.11] & 5/5 & 0.0625 \\
    Greedy & User-gap reduction (pp) & +6.54 & [5.82, 7.27] & 5/5 & 0.0625 \\
    PPO+MLP & Access success (pp) & +9.00 & [7.94, 10.06] & 5/5 & 0.0625 \\
    PPO+MLP & Mean-delay reduction & +0.24 & [0.17, 0.30] & 5/5 & 0.0625 \\
    PPO+MLP & User-gap reduction (pp) & +6.60 & [4.87, 8.33] & 5/5 & 0.0625 \\
    \bottomrule
  \end{tabular}}
\end{table}

\subsection{Behaviour Under Load}
\begin{figure}[t]
  \centering
  \begin{tikzpicture}
  \begin{axis}[
      width=\columnwidth, height=5.6cm, ybar, bar width=7pt,
      xlabel={Primary-user load}, ylabel={Packet-present access success (\%)},
      symbolic x coords={Normal, High, Extreme}, xtick=data,
    legend style={font=\scriptsize, at={(0.5,1.03)}, anchor=south,
                     legend columns=3, draw=none},
      ymin=70, ymax=100, grid=major,
    ]
    \addplot+[ybar,fill=sdnblue,draw=black!60,error bars/.cd,y dir=both,y explicit] coordinates {(Normal,93.283) +- (0,0.277) (High,89.079) +- (0,0.385) (Extreme,81.289) +- (0,1.147)};
    \addlegendentry{Greedy}
    \addplot+[ybar,fill=ppocolor,draw=black!60,error bars/.cd,y dir=both,y explicit] coordinates {(Normal,87.666) +- (0,0.804) (High,82.410) +- (0,1.224) (Extreme,74.589) +- (0,2.377)};
    \addlegendentry{PPO+MLP}
    \addplot+[ybar,fill=attention,draw=black!60,error bars/.cd,y dir=both,y explicit] coordinates {(Normal,93.853) +- (0,0.235) (High,92.298) +- (0,0.523) (Extreme,88.964) +- (0,0.556)};
    \addlegendentry{TACAN}
  \end{axis}
\end{tikzpicture}
  \caption{Packet-present access success by primary-user load, mean
    $\pm$1 standard deviation over \NumSeeds{} seeds.}
  \label{fig:packet_access_load}
\end{figure}
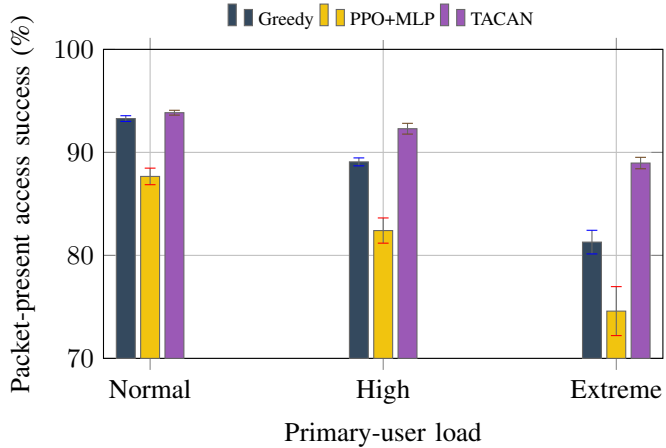

Fig.~\ref{fig:packet_access_load} and Table~\ref{tab:packet_load} decompose
packet-present access by primary-user load; SU arrivals remain fixed. TACAN and
Greedy are close at normal load
(\PacketTacanAccessNormal{}\% versus \PacketGreedyAccessNormal{}\%), a
\PacketTacanVsGreedyAccessNormalPP{}-point difference. At extreme load the
difference grows to \PacketTacanVsGreedyAccessExtremePP{} points
(\PacketTacanAccessExtreme{}\% versus \PacketGreedyAccessExtreme{}\%). This
pattern holds in every seed and is consistent with temporal context becoming
more useful as idle periods become scarce. It does not establish the mechanism
causally because the complete TACAN and Greedy configurations differ in several
ways.

% AUTO-GENERATED -- do not edit
\begin{table}[htbp]
  \centering
  \caption{Packet-present access success (\%) by primary-user load, averaged
  over five seeds.}
  \label{tab:packet_load}
  \begin{tabular}{lccc}
    \toprule
    \textbf{Method} & \textbf{Normal} & \textbf{High} &
    \textbf{Extreme} \\
    \midrule
        Greedy & 93.28 & 89.08 & 81.29 \\
    PPO+MLP & 87.67 & 82.41 & 74.59 \\
    \textbf{TACAN} & \textbf{93.85} & \textbf{92.30} & \textbf{88.96} \\
    Current-slot genie & 90.34 & 87.04 & 82.69 \\
    \bottomrule
  \end{tabular}
\end{table}

\subsection{Per-Class Service and Fairness}
Table~\ref{tab:packet_class} shows that TACAN's packet-present access success is
similar across classes: \PacketTacanUrllcAccess{}\% for URLLC,
\PacketTacanMmtcAccess{}\% for mMTC and \PacketTacanEmbbAccess{}\% for eMBB.
The corresponding Greedy values are \PacketGreedyUrllcAccess{}\%,
\PacketGreedyMmtcAccess{}\% and \PacketGreedyEmbbAccess{}\%. Mean delivery
delays are also lower with TACAN in all three classes. The URLLC miss rate is
\PacketTacanUrllcDeadline{}\% for TACAN and \PacketGreedyUrllcDeadline{}\% for
Greedy. Because the class weights, deadline penalty and QoS token are not
ablated separately, these differences cannot be assigned to one component.

% AUTO-GENERATED -- do not edit
\begin{table}[htbp]
  \centering
  \caption{Packet-present access success and mean delivery delay by traffic
  class, averaged over five seeds.}
  \label{tab:packet_class}
  \resizebox{\columnwidth}{!}{%
  \begin{tabular}{lcccccc}
    \toprule
      & \multicolumn{2}{c}{\textbf{URLLC}} &
        \multicolumn{2}{c}{\textbf{mMTC}} &
        \multicolumn{2}{c}{\textbf{eMBB}} \\
    \cmidrule(lr){2-3}\cmidrule(lr){4-5}\cmidrule(lr){6-7}
    \textbf{Method} & \textbf{Access (\%)} & \textbf{Delay} &
    \textbf{Access (\%)} & \textbf{Delay} &
    \textbf{Access (\%)} & \textbf{Delay} \\
    \midrule
        Greedy & 89.72 & 1.209 & 90.01 & 1.211 & 89.99 & 1.196 \\
    PPO+MLP & 83.14 & 1.353 & 83.43 & 1.367 & 84.26 & 1.339 \\
    \textbf{TACAN} & \textbf{92.56} & \textbf{1.121} & \textbf{92.53} & \textbf{1.124} & \textbf{92.50} & \textbf{1.121} \\
    \bottomrule
  \end{tabular}}
\end{table}

Table~\ref{tab:packet_fairness} separates two fairness questions. TACAN reduces
the max--min conditional-access gap from \PacketGreedyCondGap{} to
\PacketTacanCondGap{} points and the mean-delay gap from
\PacketGreedyDelayGap{} to \PacketTacanDelayGap{} slots. In contrast, the
per-slot delivery-rate gaps are nearly identical
(\PacketGreedyDeliveryGap{} versus \PacketTacanDeliveryGap{} points) because
delivery is limited by the common arrival process. We therefore claim improved
reliability and delay balance, not improved delivered-throughput fairness.
Sequential masking prevents same-slot duplicate channel choices for both TACAN
and Greedy; their fairness difference is not caused by SU self-collisions.

% AUTO-GENERATED -- do not edit
\begin{table}[htbp]
  \centering
  \caption{Packet-aware inter-user fairness. Gaps are max--min differences
  averaged over five seeds.}
  \label{tab:packet_fairness}
    \resizebox{\columnwidth}{!}{%
  \begin{tabular}{lcccc}
    \toprule
    \textbf{Method} & \textbf{Access gap (pp)} &
    \textbf{Delivery gap (pp)} & \textbf{Delay gap} &
    \textbf{Access Jain} \\
    \midrule
        Greedy & 9.69 & 0.97 & 0.312 & 0.9983 \\
    PPO+MLP & 9.75 & 0.98 & 0.350 & 0.9980 \\
    \textbf{TACAN} & \textbf{3.15} & \textbf{0.98} & \textbf{0.066} & \textbf{0.9998} \\
    \bottomrule
    \end{tabular}}
\end{table}

\subsection{What the Attention Learns}
\begin{figure}[t]
  \centering
  \begin{tikzpicture}
  \begin{axis}[
      width=\columnwidth, height=5.2cm,
      xlabel={Channel occupancy rank (0 = least occupied)},
      ylabel={Mean attention weight},
      legend style={font=\scriptsize, at={(0.5,-0.26)}, anchor=north,
                     legend columns=3, draw=none},
      grid=major, xmin=-0.5, xmax=19.5,
    ]
    \addplot[thick,color=phygreen,mark=*,mark size=1.2pt] coordinates {(0,0.09987) (1,0.06038) (2,0.03771) (3,0.02785) (4,0.02420) (5,0.02587) (6,0.03070) (7,0.04024) (8,0.04704) (9,0.05058) (10,0.05091) (11,0.05171) (12,0.05140) (13,0.05190) (14,0.05010) (15,0.05311) (16,0.05556) (17,0.05466) (18,0.05705) (19,0.07899)};
    \addlegendentry{Normal load}
    \addplot[thick,color=ppocolor,mark=*,mark size=1.2pt] coordinates {(0,0.11771) (1,0.06022) (2,0.03884) (3,0.03000) (4,0.03016) (5,0.03471) (6,0.04177) (7,0.04678) (8,0.04734) (9,0.04732) (10,0.04626) (11,0.04564) (12,0.04614) (13,0.04746) (14,0.05003) (15,0.05012) (16,0.04833) (17,0.04357) (18,0.04968) (19,0.07775)};
    \addlegendentry{High load}
    \addplot[thick,color=urllc,mark=*,mark size=1.2pt] coordinates {(0,0.14172) (1,0.06143) (2,0.03796) (3,0.03516) (4,0.03915) (5,0.04294) (6,0.04511) (7,0.04461) (8,0.04386) (9,0.04290) (10,0.04179) (11,0.04217) (12,0.04422) (13,0.04559) (14,0.04614) (15,0.04442) (16,0.03819) (17,0.03726) (18,0.04720) (19,0.07785)};
    \addlegendentry{Extreme load}
    \draw[dashed,gray] (axis cs:-0.5,0.05000) -- (axis cs:19.5,0.05000);
    \node[gray,font=\scriptsize,anchor=south west]
      at (axis cs:11.0,0.05000) {uniform ($1/N$)};
  \end{axis}
\end{tikzpicture}
  \caption{Trained attention of the QoS token over channel tokens (final block,
    averaged over heads and \NumSeeds{} seeds) against occupancy rank. The
    distribution is U-shaped and departs substantially from uniform $1/K$
    (dashed).}
  \label{fig:attention_viz}
\end{figure}
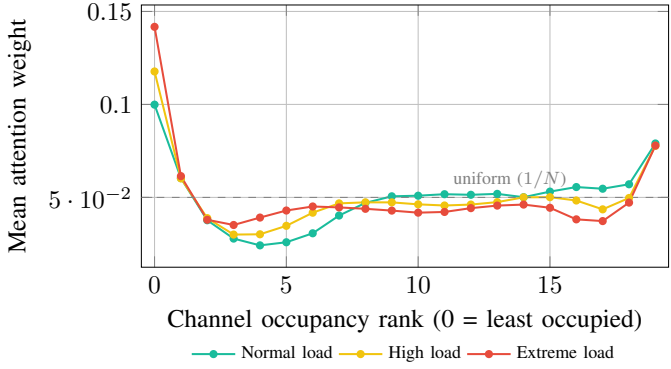

Fig.~\ref{fig:attention_viz} plots the attention of the QoS token over the
channel tokens in the final encoder block, against each channel's occupancy rank
in the current window. These are the trained weights the policy actually uses,
extracted from the encoder's self-attention rather than from a separate probe.
Channel identity is arbitrary and differs by seed, so ranking by occupancy is the
only meaningful alignment.

The distribution is non-uniform and U-shaped. Mass
concentrates on the least occupied channels, which are the transmission
candidates, \emph{and} on the most occupied ones, which are the dominant
interferers, while mid-rank channels sit at or below uniform $1/K$. The peaks
reach roughly twice uniform, rising from \AttnPeakNormal{} at nominal load to
\AttnPeakExtreme{} under extreme load. Attention to busy channels may summarize
channel persistence and overall congestion, but the weights alone do not
establish a causal mechanism. This pattern is therefore treated as descriptive.

\subsection{Ablations}
\label{sec:ablation}
Table~\ref{tab:ablation} separates two questions that are often conflated.

Group (a) compares complete encoders: TACAN has channel tokens, attention and an
auxiliary occupancy head (485,892 parameters), whereas PPO+MLP is a 155,413-
parameter flat network without the auxiliary loss or greedy-prior term. The
\TacanVsMlpPP{}-point assignment-success gap therefore compares full
configurations and does not isolate attention.

Group (b) is input perturbation: the trained policy is left untouched and one
modality is suppressed at inference. Zeroing the modulation entropy costs
\AblNoModDrop{}~points and collapsing the history to its most recent frame costs
\AblNoHistDrop{}~points in standby assignment success. These numbers must be
read with care. Zeroing drives a feature to a value the policy never observes
during training, so the drop mixes loss of information with distribution shift;
group (c) separates the two for the AMC feature. The unperturbed row reproduces
the stored assignment-success result exactly.

% AUTO-GENERATED -- do not edit
\begin{table}[htbp]
  \centering
    \caption{Standby-assignment ablation. Group (a) compares complete retrained encoders; the
    MLP is smaller and lacks TACAN's auxiliary head, so the row is not an attention-only
    ablation. Group (b) suppresses an input modality at inference time for the \emph{same}
        trained TACAN policy. $\Delta A$ is the assignment-success change in percentage points.}
  \label{tab:ablation}
  \resizebox{\columnwidth}{!}{
    \begin{tabular}{lccc}
    \toprule
    \textbf{Configuration} & \textbf{Assignment success (\%)} & \textbf{Assignment failure (\%)}
        & $\boldsymbol{\Delta A}$ \textbf{(pp)} \\
    \midrule
    \textbf{TACAN (full)} & \textbf{92.93 $\pm$ 0.14} & \textbf{7.07} & n/a \\
    \midrule
    \multicolumn{4}{l}{\emph{(a) Complete encoder comparison (independently retrained, shared PPO schedule)}} \\
    \quad PPO+MLP (155k, no auxiliary head) & 86.11 $\pm$ 0.77 & 13.89 & $-$6.82 \\
    \midrule
    \multicolumn{4}{l}{\emph{(b) Input-perturbation ablation (same trained policy, feature suppressed at inference)}} \\
    \quad Unmodified observation & 92.93 $\pm$ 0.14 & 7.07 & n/a \\
    \quad $-$ AMC-entropy feature & 90.04 $\pm$ 1.31 & 9.96 & $-$2.89 \\
    \quad $-$ Occupancy history (single frame) & 88.16 $\pm$ 0.24 & 11.84 & $-$4.77 \\
    \bottomrule
  \end{tabular}}
\end{table}

Group (c) separates information from distribution shift for the AMC feature.
Instead of zeroing it, we blend a uniform posterior into the simulated AMC output
at evaluation time. At full blending the posterior is uniform on every channel, so
the feature carries \emph{no} information about the emitter, yet it stays inside
its natural range. The standby assignment-success cost is
\AmcDestroyDrop{}~points, an order of magnitude
smaller than the \AblNoModDrop{}~points obtained by zeroing. The honest reading is
that most of the zeroing penalty is distribution shift, and that the AMC signal is
worth roughly \AmcDestroyDrop{} assignment points in this environment. Degradation is
monotone in the blending rate, so a realistic front end with partial confidence
loss would sit between these endpoints.

\IfFileExists{generated/tab_amc_robustness.tex}{%
  % AUTO-GENERATED -- do not edit
\begin{table}[htbp]
  \centering
    \caption{Standby-assignment sensitivity to AMC posterior flattening, over
  5 matched seeds. The mixing rate is the
  fraction of a uniform posterior blended into the simulated AMC output.}
  \label{tab:amc_robustness}
    \resizebox{\columnwidth}{!}{%
  \begin{tabular}{lccc}
    \toprule
    \textbf{Uniform mix (\%)} & \textbf{Assignment success (\%)} &
    $\boldsymbol{\Delta A}$ \textbf{(pp)} & \textbf{Failure (\%)} \\
    \midrule
    0 (original) & 92.93 $\pm$ 0.14 & +0.00 & 7.07 \\
    25 & 92.89 $\pm$ 0.17 & -0.04 & 7.11 \\
    50 & 92.72 $\pm$ 0.31 & -0.21 & 7.28 \\
    75 & 92.64 $\pm$ 0.35 & -0.29 & 7.36 \\
    100 & 92.59 $\pm$ 0.32 & -0.35 & 7.41 \\
    \bottomrule
    \end{tabular}}
\end{table}
}{}

\subsection{Simulation Decision-Path Cost}
\label{sec:sdnlat}
Table~\ref{tab:sdn_overhead} reports a Python microbenchmark of the
simulation-mode decision path. State extraction, policy inference and
rule-dictionary construction take \SdnTotal{}\,ms at the median and
\SdnTotalPtile{}\,ms at the 95th percentile on one CPU core. Policy inference
dominates at \SdnDecide{}\,ms. The measurement excludes sensing, AMC inference,
OpenFlow serialization and transport, switch rule installation, and radio
retuning; it is therefore not an end-to-end networking latency. With
\TacanParams{} parameters, the model occupies a few megabytes in single
precision.

% AUTO-GENERATED -- do not edit
\begin{table}[htbp]
  \centering
    \caption{Simulation-mode controller decision-path latency (1000 cycles,
    single CPU core). The measurement excludes sensing, AMC inference, OpenFlow
    transport and installation, and radio channel switching.}
  \label{tab:sdn_overhead}
  \begin{tabular}{lcc}
    \toprule
    \textbf{Component} & \textbf{Median (ms)} & \textbf{95th pct.\ (ms)} \\
    \midrule
    Simulation state extraction & 0.041 & 0.087 \\
    Policy inference & 4.943 & 17.097 \\
    Rule-dictionary construction & 0.033 & 0.069 \\
    \midrule
    \textbf{Total measured path} & \textbf{5.01} & \textbf{17.27} \\
    \bottomrule
  \end{tabular}
\end{table}

%% ====================================================================
\section{Discussion}
\label{sec:discussion}
%% ====================================================================

\subsection{Interpreting the Packet-Aware Result}
The packet-aware replay changes the meaning of the legacy result. A high standby
assignment-success rate does not imply high packet throughput because most
queues are empty under the 30\% arrival process. TACAN and Greedy therefore
deliver almost the same number of packets per SU-slot. The discriminating result
is conditional reliability: when a packet is queued, TACAN is more likely to
assign a collision-free channel, particularly under high PU load. This reduces
delivery delay and conditional reliability imbalance. It does not demonstrate
increased offered traffic or packet throughput.

\subsection{What Each Component Contributes}
The evidence supports the TACAN configuration as a package. Relative to
PPO+MLP, packet-present access improves by \PacketTacanVsMlpAccessPP{} points and
mean delay falls by \PacketTacanVsMlpDelayReduction{} slots. The comparison also
changes tokenisation, attention, auxiliary supervision, the greedy-prior term
and capacity, so it does not identify the contribution of attention alone.

The AMC-derived entropy feature contributes a small but consistent gain, and the
size of that gain depends on how it is measured. Zeroing it costs
\AblNoModDrop{} standby assignment points, but destroying its information content
while keeping it in range costs only \AmcDestroyDrop{} points, so the larger
number is mostly distribution shift rather than lost information. Even the
smaller result is not evidence that the policy identifies modulation order,
since entropy discards the posterior argmax; the simulated class templates differ
in concentration, so entropy is merely correlated with traffic class and hence
with holding time. The feature is a genuine but modest contributor here, and its
value under a real AMC front end is untested.

Packet-present reliability is similar across URLLC, mMTC and eMBB, and TACAN
improves it over Greedy in each class. The reward weights, deadline penalty and
QoS token all affect these results; their individual contributions are not
isolated.

Behaviour cloning and the learned greedy-prior term provide an engineering warm
start. No no-warm-start experiment was run, so neither convergence stability nor
final reliability is attributed to those components individually.

\subsection{Cost and Integration Scope}
The whole model is \TacanParams{} parameters, a few megabytes in single
precision, and the measured simulation decision path costs \SdnTotal{}\,ms
(Sec.~\ref{sec:sdnlat}). Attention over $K$ channel tokens is
$O(K^2 d)$, which at $K=\NumChannels{}$ is negligible; the quadratic term would
begin to matter only for wideband deployments with hundreds of channels, where
the standard remedies of local windows or channel grouping apply directly
because the token axis is the frequency axis. Nothing here requires a GPU at
inference.

\subsection{Limitations}
\label{sec:limits}
Five limitations bound the result. First, the policies were trained with
continuous per-slot assignments, including standby assignments for empty queues;
packet-aware metrics are a post hoc replay rather than packet-gated retraining.
Second, only five trained seeds are available. Parametric intervals are reported,
but the exact sign test cannot reject zero at 5\% with five all-positive pairs.
Third, PPO+MLP is not capacity- or supervision-matched to TACAN. Fourth, sensing
is ideal and AMC posteriors come from synthetic class-conditioned templates;
only one channel/user configuration is trained. Fifth, evaluation is entirely
simulated, and the controller timing excludes the live networking path. These
limitations preclude claims of attention causality, packet-throughput gain,
general scaling or deployed SDN performance.

%% ====================================================================
\section{Conclusion}
\label{sec:conclusion}
%% ====================================================================

TACAN treats the spectrum as a token sequence rather than a flat vector, gives
each channel token an AMC-derived entropy feature alongside its occupancy
history, and lets a QoS context token query the resulting representation so one
policy adapts to queue context. A packet-aware replay of the five frozen
checkpoints yields \PacketTacanAccess{}\% packet-present access success, compared
with \PacketGreedyAccess{}\% for Greedy and \PacketMlpAccess{}\% for PPO+MLP.
TACAN also lowers mean delivery delay to \PacketTacanDelayMean{} slots and
reduces the conditional user-reliability gap to \PacketTacanCondGap{} points.
The gain over Greedy increases with primary-user load and appears in all five
seeds. Packet deliveries per SU-slot remain essentially unchanged because the
experiment is arrival-limited, so no packet-throughput gain is claimed. The
result supports the complete TACAN configuration as a reliability-oriented
channel-assignment policy within the evaluated simulator. Section~\ref{sec:limits}
summarizes the scope of that evidence.

\bibliographystyle{IEEEtran}
\bibliography{paper}

\end{document}